# Optimizing the Bent Crystal Parameters for High-Efficiency Beam Extraction and Collimation in Circular Accelerators

I. A. Yazynin, V. A. Maisheev, and Yu. A. Chesnokov
Institute for High Energy Physics, st. Pobedy 1, Protvino, Moscow region, 142281 Russia

## Abstract

The efficiency of the beam extraction and collimation systems in circular accelerators with the use of the channeling effect in a bent crystal is determined. The dependences of the extraction efficiency on the geometrical parameters of the crystal (its length, thickness, and bending angle), the azimuthal location of the system components, and the offset of the septum or scraper are presented. The influence of crystal imperfections (amorphous layers, misorientation, and torsion) on the efficiency of the systems is considered, and their tolerances are proposed. It is shown that an extraction efficiency of >95% can be attained over a wide energy range from 2 GeV to 7 TeV by optimizing the crystal parameters and the positions of the system components.

## INTRODUCTION

The method for deflecting charged particles by channeling in bent crystals was proposed and theoretically justified by Tsyganov in 1976 [1]. Since that, it has been verified in many experiments (e.g., see [2,3]), which have stimulated development and investigations of crystal applications at high-energy accelerators. This method has also been implemented at the U-70 accelerator of the Institute for High Energy Physics, where crystals are used in regular runs to extract proton beams.

Development of collimation systems using crystals [6–9] is a paramount task today in view of the necessity to increase the beam intensity at the operating accelerators (in Protvino and at Fermilab) and to construct new high-energy accelerators (at CERN, in Darmstadt, and in Dubna). In this paper, we demonstrate that careful optimization of bent short crystals helps to considerably increase the efficiencies of the systems for beam extraction and collimation.

## OPERATING PRINCIPLE OF THE EXTRACTION AND COLLIMATION SYSTEMS

The typical diagram of beam extraction using the scattering target is shown in Fig. 1. The target may be either amorphous or crystalline. In this paper, we consider the case of a short crystal bent at a small angle in order to increase the amplitude of betatron oscillations so that particles can be hit to the scraper face for the collimation system or into the septum gap for the extraction system. The location of the beam and the system components in the extraction phase plane at the crystal site is shown in Fig. 2. The beam is slowly transported to the crystal via distortion of the closed orbit by the bump magnets, which allows the beam current to be adjusted for experimental setups.

Extraction efficiency $I_{out}$ is determined by the relative value of extracted protons penetrating into the septum gap. If we assume that losses of protons at the crystal $I_c$, determined by their nuclear interactions with the crystal materials, is known beforehand, the efficiency is defined by the relationship:

$$I_{out} = 1 - I_c - I_s - I_a, \qquad (1)$$

where $I_s$ is the losses at the septum partition, and $I_a$ is the number of protons scattered with large amplitudes. The efficiency of the collimation system is governed by the scattered protons that escape from the system and are lost at the vacuum chamber of the accelerator. Let us define the fraction of such protons relative to all intercepted protons as global losses $I_g$. In our case, there are only two main sources of losses — the crystal and the scraper:

$$I_g = I_{gc} + I_{gs}. \qquad (2)$$

Should the nuclear loss of protons at the crystal be known, the global loss due to the crystal can be estimated as $I_{gc} = I_c \kappa_1$, where the loss due to the scraper is defined by particle density $\rho_s$ [mm⁻¹] at the scraper edge $I_{gs} = \rho_s \kappa_2$. Coefficients $\kappa_1$ and $\kappa_2$ are determined by the position of the scraper and the collimators.

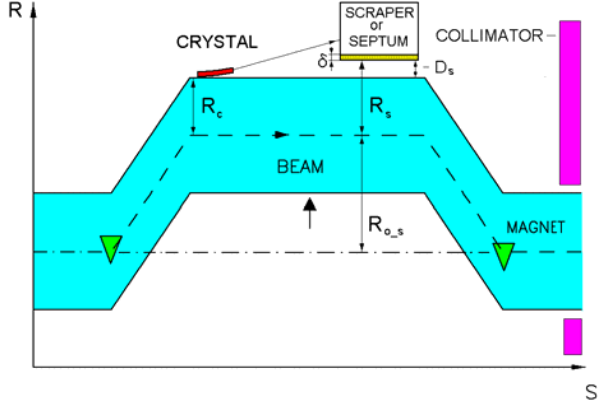
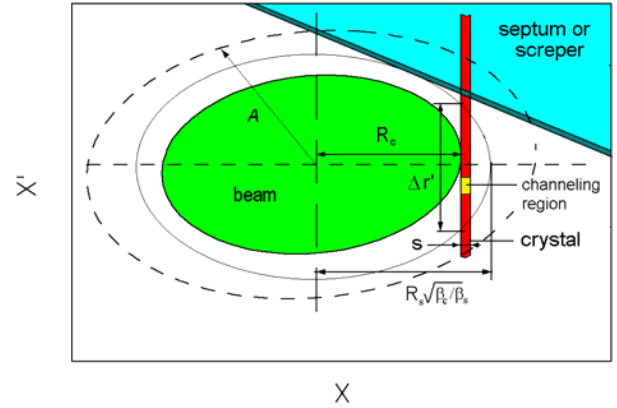

Fig. 1. Schematic diagram of the extraction system.   Fig. 2. Phase plane of the beam at the crystal site.

Computer simulation of the beam interception process shows that, for the beam forming system at the U-70 accelerator, $\kappa_1 \approx 0.05$ and $\kappa_2 \approx 0.08$. Therefore, the efficiencies of beam extraction and beam halo interception increase as the beam loss at the crystal and the proton density at the scraper or septum edge decrease.

The efficiencies of beam extraction and collimation depend on the crystal dimensions and imperfection, as well as on the position of the crystal and the septum in the accelerator. The necessary value of the beam casting by the crystal is determined by its azimuthal position and the septum (scraper) offset, which is defined as the distance between it and the circulating beam (Fig. 1):

$$D_s = R_s - R_c \sqrt{\beta_s / \beta_c} = |R_s| - R_{s0}, \qquad (3)$$

where $R_s$, $R_c$ are the septum and crystal positions with respect to the equilibrium orbit in the extraction plane, $\beta_s$ and $\beta_c$ are the respective amplitude (beta) functions, and $R_{s0}$ is the half-size of the circulating beam at the septum site. The minimum crystal bending angle sufficient to get the beam behind the septum partition with thickness $\delta$ is determined from the equation of motion:

$$\Delta x' = \vartheta + [(R_s + \delta) - R_c \cdot m_{11}]/m_{12}, \qquad (4)$$

where $m_{11} = \sqrt{\beta_s / \beta_c} \cos \Delta\psi$, $m_{12} = \sqrt{\beta_s \beta_c} \sin \Delta\psi$ are the elements of the transfer matrix; $\Delta\psi = \psi_s - \psi_c$ is the phase shift of betatron oscillations from the crystal to the scraper; and $\vartheta$ is the angular half-size of the beam emerging from the crystal, which corresponds to the critical angle for the perfect crystal. Based on the known angular density of particles scattered by the crystal $P_c(\Delta x_c')$, it is possible to determine the density at the septum partition: $\rho_s(R_s) = P_c(\Delta x_{cs}')/m_{12}$. From these dependences, it follows that, for the length of the crystal (and the loss at it) and the loss at the septum partition to be reduced, matrix element must be rather great; i.e., the optimum phase shift is in the range π/3 + πn < Δψ < π/2 + πn. In the U-70 accelerator, there are two magnets between the crystal and the septum (the scraper), which corresponds to phase shift $\Delta\psi \approx \pi/3$; therefore, their position is nearly optimal.

## CALCULATION OF THE SYSTEMS FOR BEAM EXTRACTION AND COLLIMATION AT THE U-70 ACCELERATOR

Let us investigate the effect of septum offset on the beam extraction efficiency, which is determined by computer simulation of the 50-GeV proton beam interception process. The beam is simultaneously guided by the bump magnets to the crystal and the scraper located in gaps 84 and 86, respectively. All of the above calculations of the beam extraction and collimation systems were performed using the SCRAPER code [5], which allows one to simulate beam dynamics in actual magnetic fields of the accelerator, taking into account interaction of particles with the crystal and the other components of the systems.

Let us consider protons get beyond the septum partition or the scraper edge with thickness $\delta = 1$ mm (Fig. 1) as normalized intensity of the extracted beam $I_{out}$ or the extraction efficiency. By default, a short perfect silicon crystal with orientation (110), length downstream of the beam $L = 0.9$ mm, bending angle $\alpha = 1.1$ mrad, and thickness in the lateral plane $s = 1.4$ mm was used in the calculations. Figure 3 presents the dependences of the extraction efficiency on the angular position of the crystal (orientation curves) at different scraper offsets.

At the optimal angular position, extraction efficiency $I_{out}$ and loss at the scraper $I_s$ depend on the offset (Fig. 4). It is apparent that the efficiency does not grow in value at offsets $D_s > 2$ mm, but the region of high efficiency adjustment becomes wider.

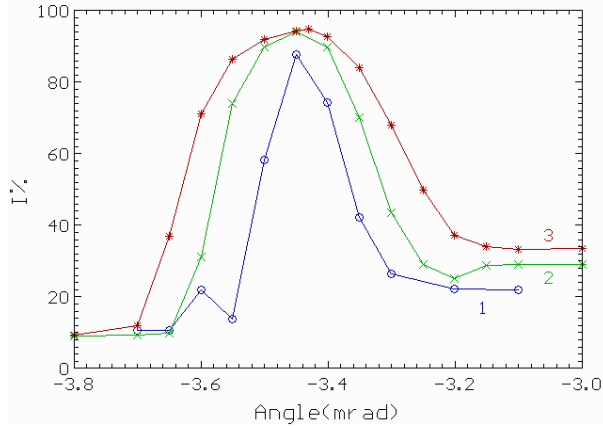 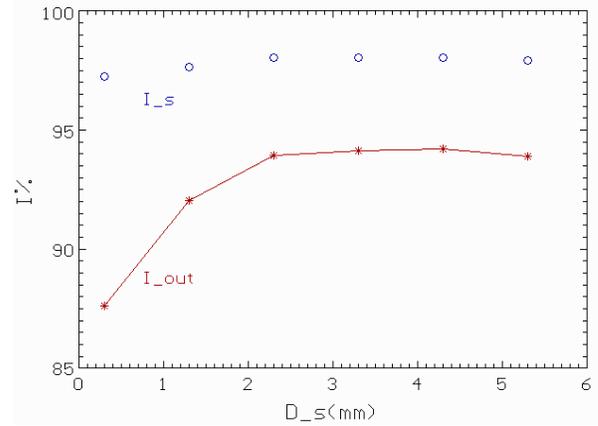

Fig. 3. Orientation curves for offsets (*1*) of 0.3, (*2*) 2.3, and (*3*) 4.0 mm.

Fig. 4. Dependence of the extraction efficiency on the offset value.

At small offsets, a major portion of scattered particles that have not been involved in the channeling mode are immediately incident on the scraper edge, thus increasing the losses. The widening of the high-efficiency region from the offset is explained by the increased width of the crystal's angular adjustment region, in which, immediately or after being scattered, particles of the circulating beam can be trapped into the channeling mode and be extracted to the scraper (Fig. 2):

$$\Delta r' = 2\sqrt{R_s^2/\beta_s\beta_c - R_c^2/\beta_c^2} = 2\sqrt{(D^2 + 2DR_{s0})/\beta_s\beta_c} \ . \qquad (5)$$

When the beam is guided, the offset value and the width of the region depend on the circulating beam size, which varies from the maximum $R_c = 4.5$ mm to zero. In the case of beam halo interception (this happens in localization of beam losses in colliders), these parameters remain constant, since the beam size does not change. In the subsequent calculations, we selected offset $D_s$

= 4 mm at proton amplitude $R_{s0}$ = 4 mm, which corresponds to the width of the angular adjustment region $\Delta r'$ = 0.6 mrad (see Fig. 3). At very large offsets, beam casting must be intensified, which leads to an increase in the crystal length and, hence, in the loss at it.

The extraction efficiency is also affected by the speed of beam guidance to the crystal. The calculated orientation curves of the extraction efficiency and the respective losses at the crystal at different speeds are presented in Fig. 5. At high guidance speeds, the extraction efficiency and the width of the high efficiency region decrease, which can be attributed to the fast beam passage through the crystal, and a significant portion of protons have no time to enter the channeling mode.

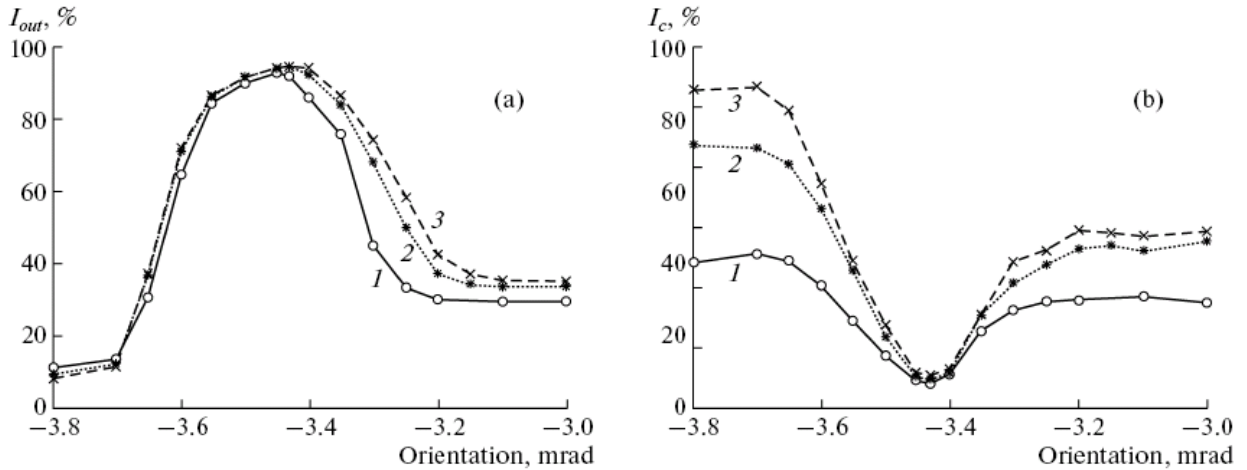

Fig. 5. Orientation curves (a) of the extraction efficiency and (b) loss at the crystal at guidance speeds of (*1*) 0.92273, (*2*) 0.00568, and (*3*) 0.00227 mm/turn.

The decrease in the proton intersection multiplicity causes the loss at the crystal to decrease (Fig. 5b). At optimum crystal position, the efficiency starts lowering at a guidance speed of >0.01 mm/turn and decreases from 95 to 92% at a guidance speed of 0.02 mm/turn. Further calculations were performed at a guidance speed of 0.005 mm/turn; i.e., the beam with a radius of 4.5 mm was fully intercepted over 1000 turns. In this case, the loss at the crystal decreases by a factor of ~10 in the optimum position in comparison with the case of disorientation.

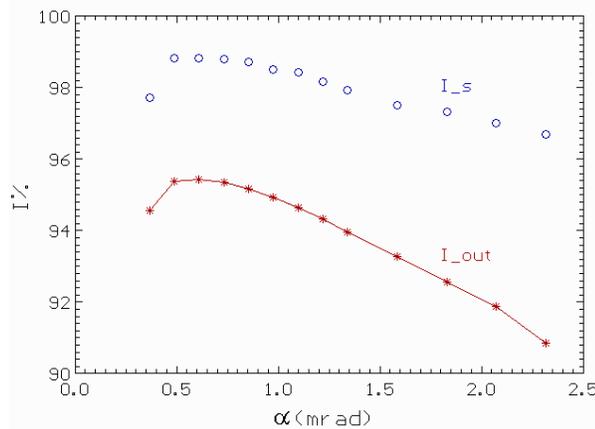

Fig. 6. Dependence of the extraction efficiency on the crystal bending angle at constant curvature radius R = 0.82 m.

The extraction efficiency is also affected by the crystal bending angle (Fig. 6), which corresponds to the angle of deflection of the channeling particles and, hence, their casting to the scraper (the septum). For protons to be get to the scraper edge, the deflection angle must be 0.22 mrad. At small bending angles of the crystal, a fraction of the beam is incident on the scraper edge.

At large bending angles, the crystal length increases, and increased Coulomb scattering and dechanneling cause the extraction efficiency to decline. From the efficiency dependence presented in Fig. 6, it follows that the optimum bending angle α is in the range of 0.5–0.8 mrad.

It should be noted that crystal imperfections cause the extracted beam to grow in size. It should also be taken into account that the region of admissible crystal alignment widens with an increase in the bending angle. Therefore, for the extraction system of a 50 GeV beam, the optimum deflection angle grows in value to 0.7–0.8 mrad, and the extraction efficiency becomes ~95%.

The influence of the curvature radius on the extraction efficiency is determined at the next step. Figure 7a presents the extraction efficiency and losses at the scraper as functions of crystal length $L$ at constant bending of the angle of a silicon crystal α = 1.1 mrad.

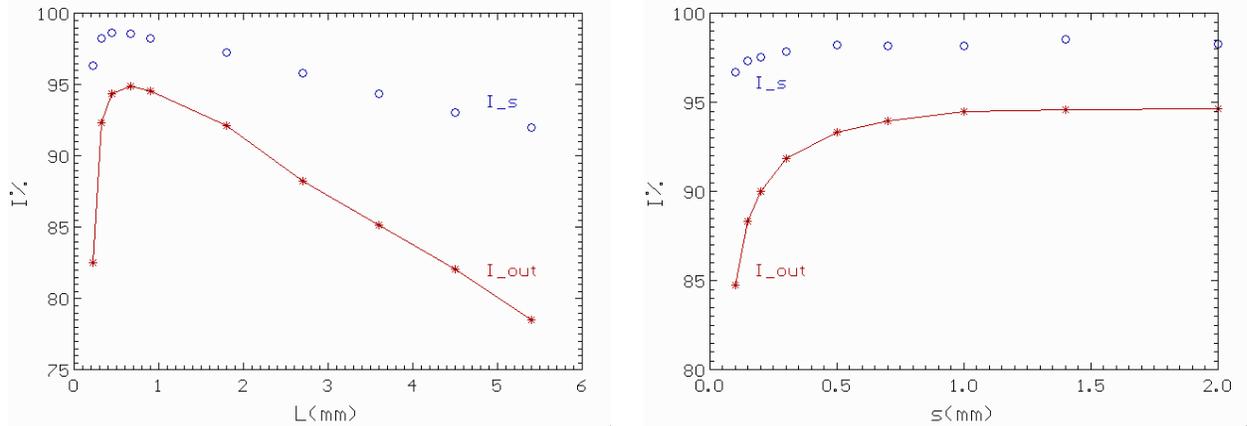

Fig. 7. Dependences of the extraction efficiency (a) on the crystal length and (b) thickness.

For very short crystals, curvature radius $R = L/\alpha$ is small, and the losses increase through higher dechanneling. As noted above, they will also increase at greater lengths. The maximum efficiency will be obtained at crystal length $L = 0.7$–$0.8$ mm, which corresponds to curvature radius $R = 0.6$–$0.7$ m. The calculations for other accelerators with different energies show that the optimum curvature radius is 7–9 critical radii $R_{crit}$.

The extraction efficiency grows in value with an increase in the crystal thickness $s$ in the transverse plane (Fig. 7b). For very thin crystals, initially scattered particles that avoided the channeling mode enter repeatedly the region of trapping into the channeling mode with a lower probability (Fig. 2). At crystal thickness of >1 mm, the extraction efficiency becomes as high as ~95% and remains constant.

For the short crystals under investigation, the extraction efficiency increases with amorphous layer thickness $s_{am}$ only slightly owing to the increase in the number of proton intersections with the crystal (Fig. 8a). In thicker layers ($s_{am}$ ~ 100 μm), protons initially interact with the crystal as with an ordinary scattering target and increase the amplitude of betatron oscillations. In subsequent intersections, the protons are incident mainly on the region of the perfect crystal. In other words, the mean number of intersections increases by factors of 1–2, and nuclear interactions cause the loss at the crystal to increase by value $\Delta I_c = \Delta N L / L_n = 0.2$–$0.4$%, where $L_n$ is the nuclear length of silicon. The extraction efficiency decreases approximately by this value.

In the case of crystal torsion (deviation from the cylindrical bending), the angular orientation of the input nuclear planes of the crystal depends on vertical coordinate $z$: $\Delta\theta = zT$. If the vertical half-size of the beam is 4 mm and the torsion parameter is rather high ($T = 0.02$ mrad/mm), the maximum deflection is $\Delta\theta_{max} = \pm 0.08$ mrad, which leads to a small decrease in the extraction efficiency (~1%) and an increase in the radial beam size at the absorber. For comparison, Fig. 8b presents the orientation curves of the efficiency for a perfect crystal and a crystal with torsion. One can observe the widened region of admissible crystal alignment due to the presence of torsion.

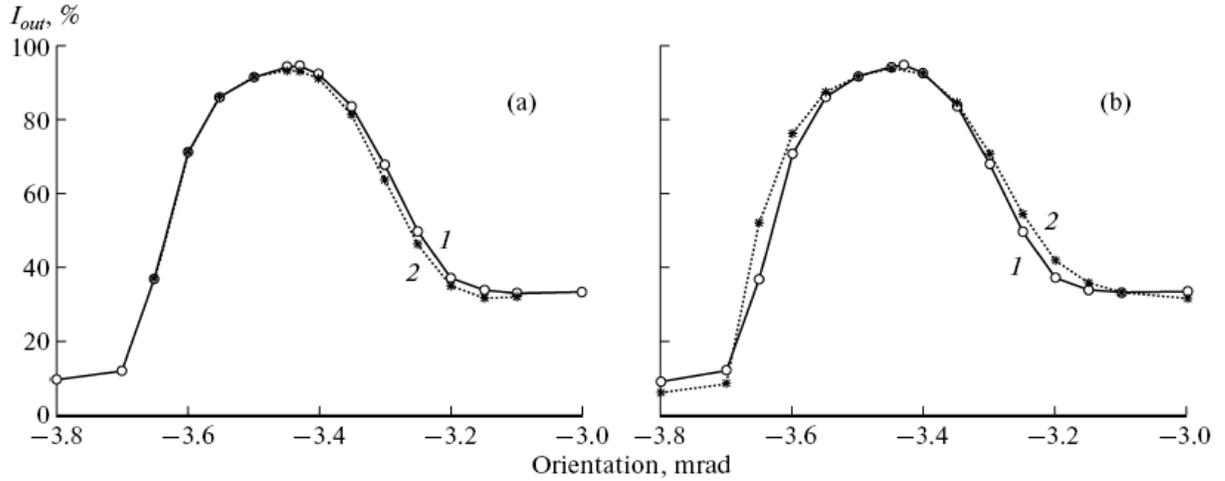

Fig. 8. Dependences of the extraction efficiency (a) on the amorphous layer thickness (**1**) $s_{am} = 0$ and (**2**) 100 μm, and (b) on the torsion parameter of the crystal (**1**) $T = 0$ and (**2**) 0.002 mrad/mm.

The other crystal imperfections, such as the misorientation (the angle between the face and the crystallographic plane), affect the extraction efficiency in the U-70 accelerator only slightly, since the value of particle casting onto the scraper upon guiding is ~ 50 μm, which is considerably greater than the coverage of these effects. The defects in treatment of the front and back surfaces of the crystal, which are ~50 μm, reduce the value of casting by ~5–10% and the extraction efficiency by ~1%; in addition, they increase the beam size at the scraper.

The extraction efficiency can be increased to 98% by organizing optimum beam guiding to the scraper (the septum). In the operating setup described in this paper, when the beam is guided by the bump magnets, particles with different amplitudes are incident on the crystal end surface with an angular divergence of ≈0.15 mrad, which is four times as great as the area of trapping into the channeling mode. This means that, even at optimal crystal alignment, a large fraction of particles is trapped into the channeling mode repeatedly after being scattered from the crystal, thus increasing the losses at it. As a result, the losses at the crystal increase several times, up to 1%. The theoretical values of the extraction efficiency are corroborated by the experiments performed with bent crystals at the U-70 accelerator in [3, 10].

**INFLUENCE OF THE BEAM ENERGY ON THE EXTRACTION EFFICIENCY**

The crystal-based collimation systems have recently been investigated at accelerators differing in the proton beam energy and the magnetic structure at the site of the system components [3, 6–10]. Let us consider the effect that the main beam parameters (the energy and the size) exert on the extraction efficiency. For the sufficient casting of the beam onto the scraper or the septum in view of their offset from the circulating beam, the crystal must provide the deflection angle that can be determined from Eqs. (3) and (4):

$$\alpha = k\sigma / \beta = k\sqrt{\varepsilon_0 / PV\beta}, \qquad (6)$$

where parameter $k = 10–15$; $\varepsilon_0$ is the normalized emittance at the level of the rms beam size σ, $P$ and $V$ are the proton momentum and velocity, and $\beta = \sqrt{\beta_s \beta_c}$. As was noted above, the optimum crystal radius can be presented in terms of interplanar spacing $d$ and the potential energy:

$R_{opt} = 1.5dPV/eU_{max} \approx 8R_{crit}$. For a silicon crystal with the (110) orientation and the potential energy of its nuclear planes $eU_{max}$ = 21.4 eV, the optimum radius and length of the crystal are

$$R_{opt} = 0.0135PV \quad \text{and} \quad L = \alpha R_{opt} = 1.16\sqrt{\varepsilon_0 PV/\beta}. \tag{7}$$

The dependences of the channeling fraction intensity at the output of the optimal crystal on the momentum of a proton in its single passage through the crystal are presented in Fig. 9a at $k$ = 10 and 15. Channeling fraction intensity $I_{1ch}$ varies from 70 to 82% with increasing energy, which can be attributed to the reduction of dechanneling by the decreased scattering. In the calculations, we used characteristic normalized emittance $\varepsilon_0 = 3$ (mm mrad) and amplitude (beta) function β = 100 m.

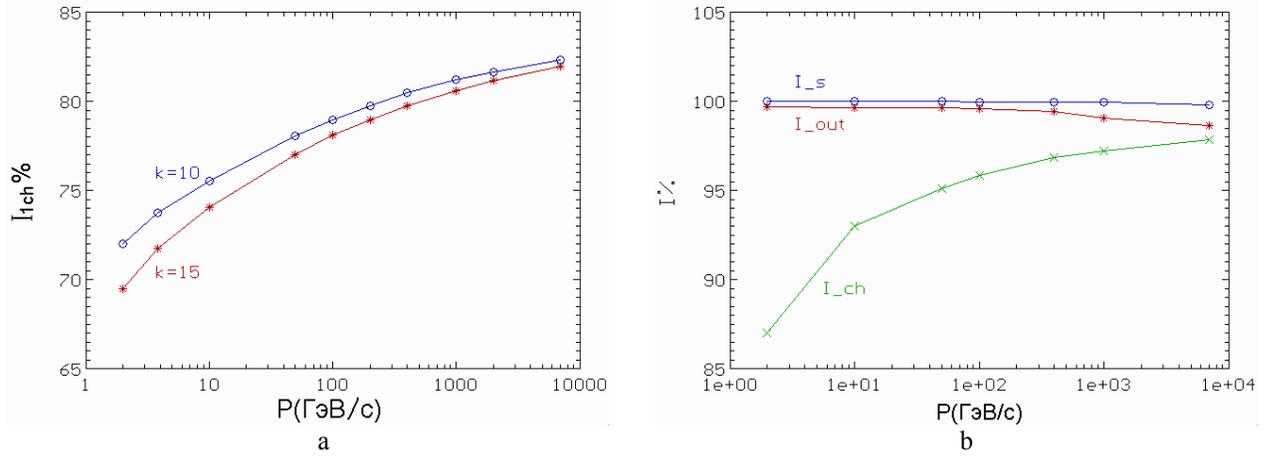

Fig. 9. Dependence of the channeling fraction intensity on the proton momentum in the cases (a) of single deflection and (b) slow multiturn extraction.

During slow extraction implying multiple interactions with the crystal, a significant fraction of the initially scattered beam is trapped into the channeling mode after several turns. Therefore, the intensity of the channeling fraction being extracted is $I_{ch} = I_{1ch} + (1 - I_{1ch}) \cdot \eta$, where the probability of repeated trapping into channeling is in the range $\eta$ = 0.7–0.9 and increases with increasing energy owing to the lower particle scattering from the crystal. As a result, the intensity of the channeling fraction (Fig. 9b) for the slow extraction increases to 87% for low energies ($P$ = 2 GeV/$c$) and to 98% for the maximum proton energy ($P$ = 7 TeV/$c$ for the Large Hadron Collider). Apart from the channeling fraction, protons from the other fractions (mainly from the dechanneling fraction) also penetrate into the septum gap when the crystal is in the optimum position. As a result, extraction efficiency may be as high as 99%. Almost all protons reach the septum, except for those suffering nuclear interactions with the crystal $I_c$ = 100% − $I_{out}$ = 0.05–0.25%. The losses at the septum partition (δ = 1 mm in the calculations) can be found from the two dependences $I_{sep} = I_s - I_{out}$; its value is 0.3% at low and medium energies and ∼1% at high energies.

The calculations have demonstrated that, when proton channeling is used in beam collimating systems, the proton density at the collimator edge is $\rho_s$ = 0.003–0.010 mm$^{-1}$, which is two orders of magnitude lower relative to traditional systems with a scattering target [4]. The losses at the crystal are also one or two orders of magnitude lower than the losses at the target. Therefore, based on Eq. (2), we can conclude that the use of crystals increases the particle interception efficiency by factors of 20–50.

As was shown earlier (see Eq. (4)) density on the septum partition is $\rho_s(R_s) \sim 1/m_{12}$, therefore, value of losses on the septum will be inversely proportional of the amplitude function $I_{sep} \sim 1/\beta$. At the expense of crystal length decrease, the value of losses on it will be $I_c \sim 1/\sqrt{\beta}$. In special technological places it is possible considerably to increase the amplitude function in 2-7 times and practically in as much time to raise efficiency of systems.

As follows from Eq. (7), the optimum crystal length increases with increasing particle energy as the square root of the energy. At low energies, the optimum crystals are short. Even at an energy of 400 GeV (SPS, CERN), the length of the optimum crystal length will be 0.8 mm, and for the energy of 1 TeV (Tevatron, Fermilab), it will be 1.3 mm. From dependence (3), it also follows that, the greater the amplitude function and the smaller the beam emittance, the shorter the crystal, and the higher the extraction efficiency. For a perfect optimum crystal, the extraction efficiency is ~99% at energies of 10–1000 GeV and 98% at 7000 GeV. The lowering of the efficiency at high energies is explained by the increase in the nuclear losses caused by the increase in the crystal length.

In conclusion, based on the calculations presented in this paper, we infer that the geometrical parameters of the crystal (its length, thickness, and bending angle) have a stronger effect on the efficiency of the extraction and collimation systems than the crystal imperfections, such as the amorphous layer, misorientation, and torsion. Using the above recommendations, it is possible to improve crystal-based systems of beam extraction and collimation at many accelerators, in particular, at the Large Hadron Collider.

## ACKNOWLEDGMENTS

This work was supported by the Russian Foundation for Basic Research, grant no. 08_02_01453_a.